\documentclass[preprint]{aastex}

\usepackage{natbib}
\def\degree{\ifmmode {^\circ}\else {$^\circ$}\fi}
\def\rstar{\ifmmode {\, R_{\star}}\else $R_{\star}$\fi}
\def\msol{\ifmmode {\, M_{\odot}}\else $M_{\odot}$\fi}
\def\rsol{\ifmmode {\, R_{\odot}}\else $R_{\odot}$\fi}
\def\lsol{\ifmmode {\, L_{\odot}}\else $L_{\odot}$\fi}
\def\msolyr{\ifmmode {\, M_{\odot}\,{\rm yr}^{-1}}\else $M_{\odot}\,{\rm yr}^{-1}$\fi}
\def\mdot{\ifmmode {\,\dot{M}}\else $\dot{M}$\fi}
\def\mdotyr{\ifmmode {\,\dot{M}\,yr^{-1}}\else $\dot{M}\,yr^{-1}$\fi}

\begin{document}

\title{The Dust Properties of Two Hot R Coronae Borealis Stars and a Wolf-Rayet Central Star of a Planetary Nebula: in Search of a Possible Link}

\author{ Geoffrey C. Clayton\altaffilmark{1}, O. De Marco\altaffilmark{2},  B. A. Whitney\altaffilmark{3,4}, B. Babler\altaffilmark{4}, J. S. Gallagher\altaffilmark{1,5}, J. Nordhaus\altaffilmark{6}, A.K. Speck\altaffilmark{7}, M.J. Wolff\altaffilmark{3}, W.R. Freeman\altaffilmark{1}, K.A. Camp\altaffilmark{1}, W.A. Lawson\altaffilmark{8}, J. Roman-Duval\altaffilmark{9}, K. A. Misselt\altaffilmark{10},  M. Meade\altaffilmark{4}, G. Sonneborn\altaffilmark{11}, M. Matsuura\altaffilmark{12,13}, and M. Meixner\altaffilmark{9}}

\altaffiltext{1}{Department of Physics \& Astronomy, Louisiana State
University, Baton Rouge, LA 70803; gclayton@fenway.phys.lsu.edu,  wfreem2@lsu.edu, kcamp5@tigers.lsu.edu}

\altaffiltext{2}{Department of Physics, Macquarie University, Sydney, NSW 2109, Australia; orsola@science.mq.edu.au}

\altaffiltext{3}{Space Science Institute, 4750 Walnut St. Suite 205, Boulder, CO
80301, USA; bwhitney, mjwolff@spacescience.org}

\altaffiltext{4}{Department of Astronomy, 475 North Charter St., University of
Wisconsin, Madison, WI 53706, USA; brian, meade@astro.wisc.edu}

\altaffiltext{5}{Department of Mathematics, Physics, \& Computer Science, Raymond Walters College,
University of Cincinnati, Blue Ash, OH 45236; gallagjl@ucmail.uc.edu}

\altaffiltext{6}{Department of Astrophysical Sciences, 116 Peyton Hall, Princeton University, Princeton, NJ 08544; nordhaus@astro.princeton.edu}

\altaffiltext{7}{Physics \& Astronomy, University of Missouri, Columbia, MO 65211; speckan@missouri.edu}

\altaffiltext{8}{School of PEMS, University of New South Wales, Australian Defence Force Academy, P.O. Box 7916, Canberra 2610, Australia; w.lawson@adfa.edu.au}

\altaffiltext{9}{STScI, 3700 San Martin Dr., Baltimore, MD 21218; duval, meixner@stsci.edu}

\altaffiltext{10}{Steward Observatory, University of Arizona, 933 North Cherry
Ave., Tucson, AZ 85721, USA; misselt@as.arizona.edu}

\altaffiltext{11}{Observational Cosmology Laboratory Code 665, NASA Goddard Space Flight Center,
Greenbelt, MD 20771, USA; george.sonneborn-1@nasa.gov}


\altaffiltext{12}{Department of Physics and Astronomy, University College London, Gower Street, London WC1E 6BT, UK; mikako.matsuura@ucl.ac.uk}

\altaffiltext{13}{MSSL, University College London, Holmbury St. Mary, Dorking, Surrey RH5 6NT, UK}


\begin{abstract}

We present new Spitzer/IRS spectra of two hot R Coronae Borealis (RCB) stars, one in the Galaxy,V348 Sgr, and one lying in the Large Magellanic Cloud, HV 2671. 
These two objects may constitute a link between the RCB stars and the late Wolf-Rayet ([WCL]) class of central stars of planetary nebula (CSPNe) such as CPD --56$^{\rm o}$~8032 that has little or no hydrogen in their atmospheres.
HV 2671 and V348 Sgr are members of a rare subclass that has significantly higher effective temperatures than most RCB stars, but sharing the traits of hydrogen deficiency and dust formation that define the cooler RCB stars. The [WC] CSPNe star, CPD --56$^{\rm o}$~8032, displays evidence for dual-dust chemistry showing both PAHs and crystalline silicates in its mid-IR spectrum. HV 2671 shows 
strong PAH emission but shows no sign of having crystalline silicates. 
The spectrum of V348 Sgr is very different from those of CPD --56$^{\rm o}$~8032 and HV 2671. The PAH emission seen strongly in the other two stars is not present. Instead, the spectrum is dominated by a broad emission centered at about 8.2 \micron. This feature is not identified with either PAHs or silicates. Several other cool RCB stars, novae and post-asymptotic giant branch stars show similar features in their IR spectra. 
The mid-IR spectrum of CPD --56$^{\rm o}$~8032 shows emission features that may be associated with C$_{60}$. The other two stars do not show evidence for C$_{60}$.
The different nature of the dust around these stars does not help us in establishing further links that may indicate a common origin.

HV 2671 has also been detected by Herschel/PACS and SPIRE. V348 Sgr and CPD --56$^{\rm o}$~8032 have been detected by AKARI/FIS. These data were combined with Spitzer, IRAS, 2MASS and other photometry to produce their spectral energy distributions from the visible to the far-IR. Monte Carlo radiative transfer modeling was used to study the circumstellar dust around these stars. HV 2671 and CPD --56$^{\rm o}$~8032 require both a flared inner disk with warm dust and an extended diffuse envelope with cold dust to to fit their SEDs. The SED of V348 Sgr can be fit with a much smaller disk and envelope.
The cold dust in the extended diffuse envelopes inferred around HV 2671 and CPD --56$^{\rm o}$~8032 may consist of interstellar medium swept up during mass-loss episodes.
 
\end{abstract}



\section{Introduction}

The R Coronae Borealis (RCB) stars are an extremely small class of variables, notable for their hydrogen deficiency and large irregular brightness variations caused by circumstellar dust formation \citep{1994JApA...15...47L,1996PASP..108..225C}.
Several evolutionary scenarios have been suggested, including a merger of two white dwarfs, and a
final helium shell flash \citep[e.g.,][]{1984ApJ...277..355W, 2007ApJ...662.1220C}. 
RCBs are mostly made of helium. This fact and the recent detection of a high $^{18}$O/$^{16}$O ratio has significantly tilted the body of evidence towards a binary merger interpretation for these stars \citep{2000A&A...353..287A,2005ApJ...623L.141C,2007ApJ...662.1220C,2009ApJ...696.1733G,2010ApJ...714..144G}.

Most RCB stars resemble F- or G-type supergiants. However, there is also an even rarer subclass of the RCB stars that has significantly higher effective temperatures ($\sim$15,000-20,000 K), but shares the traits of hydrogen deficiency and dust formation \citep{2002AJ....123.3387D}. Only three such stars, V348 Sgr, MV Sgr and DY Cen are known in the Galaxy. Recently, one additional star, HV 2671, was discovered in the Large Magellanic Cloud   \citep[LMC;][]{1996ApJ...470..583A, 2002AJ....123.3387D}. 
Although these four stars are part of a subclass referred to as the hot RCB stars, they do not necessarily share a common evolution \citep{2002AJ....123.3387D}. Both DY Cen and MV Sgr have typical RCB helium abundances ($\sim$98\%), which exclude any currently known post-asymptotic giant branch (post-AGB) evolutionary models, and once again support a binary merger origin for these stars \citep{2000A&A...353..287A,2005ApJ...623L.141C,2007ApJ...662.1220C,2009ApJ...696.1733G,2010ApJ...714..144G}. 
Observations of the other two hot RCB stars, V348 Sgr and HV 2671, on the other hand, lead to stellar abundance determinations that are more in line with those of Wolf-Rayet [WC] central stars of planetary nebulae  \citep[CSPN; He=40-50\%, C=40-50\%, O=5-10\%, by mass;][]{2000A&A...360..952H,2001Ap&SS.275...53D,2002AJ....123.3387D}. 

While the abundances of V348 Sgr have been determined by \citet{1994A&A...283..567L}, the abundances of HV 2671 have not, but its low-resolution spectrum is almost identical to that of V348 Sgr. 
The visible spectra of V348 Sgr and HV 2671, as well as their abundances, are similar to those of the [WC10] CPD --56$^{\rm o}$~8032 \citep{1998MNRAS.296..419D,2002AJ....123.3387D}. However, both V348 Sgr and HV 2671 are cooler ($\sim$20,000 K) than the [WC10]  spectral class ($\sim$34,000 K)  \citep{1994A&A...283..567L,1998MNRAS.296..419D}, such that \citet{1998MNRAS.296..367C} suggested that V348~Sgr fails the late Wolf-Rayet ([WCL]) criterion altogether. Wolf-Rayet CSPN have been explained as stars such as V605~Aql that suffer a last helium shell flash just before or after the departure of the stars from the AGB \citep[e.g.,][]{2001ApJ...554L..71H,2006ApJ...646L..69C}. However, several of their observed characteristics are not explained by this model and this has invited alternative, binary-based scenarios \citep[e.g.,][]{2002PASP..114..602D,2008ASPC..391..209D}

All of the stars are hydrogen deficient but the amount of hydrogen present varies greatly from star to star \citep{2002AJ....123.3387D}. V348 Sgr  has a measured hydrogen abundance of 4\% by mass but this may be an upper limit. HV 2671 has no measurement of its hydrogen abundance but its spectrum looks very similar to that of V348 Sgr. The hydrogen abundance by mass in  CPD --56$^{\rm o}$~8032 is $<$1\%~ \citep{1998MNRAS.296..419D}.
The hot RCB stars, V348 Sgr and HV 2671, as well as the [WC10] CPD --56$^{\rm o}$~8032 are surrounded by hydrogen-rich PNe: there is a large PN ($\sim$30\arcsec) observed around V348 Sgr \citep{1991MNRAS.248P...1P}, and a young, dense PN is seen around CPD --56$^{\rm o}$~8032 \citep{1997MNRAS.292...86D}. HV 2671 likely has a PN also \citep{2002AJ....123.3387D}, as can be inferred by its Balmer and [O III] emission lines. However, HV 2671 shows no sign of extended H$\alpha$ emission on images obtained for the Reid-Parker LMC PN survey \citep{2006MNRAS.365..401R,2006MNRAS.373..521R}, likely due to its large distance.

The lightcurve behavior of HV 2671 and V348 Sgr is typical for RCB stars with periods of inactivity followed by sudden declines in brightness by several magnitudes and slow returns to maximum light \citep{1996ApJ...470..583A,2001ApJ...554..298A,2002AJ....123.3387D,2009AcA....59..335S}. V348 Sgr, in particular, is very active, showing many declines of 4 mag or more. The declines seen in CPD --56$^{\rm o}$~8032 are similar in form to those seen in RCB stars although they are much shallower ($\sim$1 mag) \citep{1999Obs...119...76J,2002MNRAS.332..879C}. The CPD --56$^{\rm o}$~8032 declines have been associated with an obscuring disk resolved with HST and VLTI \citep{2002ApJ...574L..83D,2006A&A...455.1009C}. No such high spatial resolution observations exist for V348 Sgr and HV 2671. 

The link between RCB and [WC] CSPN may reside in a binary interaction that generated both classes. As pointed out previously, there is now compelling evidence that the RCB stars were produced in a merger of two WDs. [WC] stars, on the other hand, have abundances consistent with post-AGB single star evolution. However, one line of evidence that undermines this interpretation derives from the observations of circumstellar dust. \citet{1989ApJ...341..246C} found that CPD --56$^{\rm o}$~8032 shows the emission features ascribed to  polycyclic aromatic hydrocarbons (PAHs) in its IR spectrum. Its PN shows C/O$>$1 \citep{1997MNRAS.292...86D}, so it was a surprise to also find that the mid-IR spectrum of CPD --56$^{\rm o}$~8032 also shows the presence of many emission features attributed to crystalline silicates. The simultaneous presence of both C-rich and O-rich chemistry and dust  \citep{1997Ap&SS.251...15B,1998A&A...331L..61W,1999ApJ...513L.135C} may point to the presence of a long lived disk which may have formed through binary interaction \citep{1998A&A...331L..61W,1999ApJ...513L.135C,2002MNRAS.332..879C} and this was corroborated by the detection of an edge-on disk around this star \citep{2002ApJ...574L..83D}. These findings are in line with a binary scenario for this object and possibly for the class of [WCL] CSPNe \citep{2001ASSL..265..157Z,2002PASP..114..602D,2008ASPC..391..209D}.
It was suggested that the dual-dust chemistry phenomenon in PNe seems showed a strong correlation with the presence of a late [WC] nucleus \citep{2002MNRAS.332..879C}, but more recent Spitzer observations of the Galactic Bulge show that dual-dust chemistry is common in all (cool and hot) [WC]-type PNe \citep{2009A&A...495L...5P, 2011MNRAS.tmp..441G}. A final thermal pulse on the AGB or hydrocarbon chemistry in an UV-irradiated, dense torus may be responsible.


In this paper, we present new Spitzer/IRS spectra of HV 2671 and V348 Sgr, which have been analyzed and compared with  that of CPD --56$^{\rm o}$~8032 to look for evidence for dual-dust chemistry that may establish further links between the hot RCB and [WC] classes that that may possibly elucidate the connection between the dust and the evolution that generated it. We also present the spectral energy distributions (SEDs) of the three stars from the visible to the far-IR using new Spitzer, AKARI and Herschel photometry. These SEDs have been modeled using a Monte Carlo radiative transfer (RT) code to determine the nature of the circumstellar dust around these stars and infer their mass-loss history.

\section{Observations and Data Reduction}

The new Spitzer/IRS spectra of HV 2671 and V348 Sgr were obtained in Stare mode as part of program 30380. 
The two targets were observed with the Short-Low (SL; $5.2-14$ $\mu$m; $\lambda/\Delta\lambda\sim90$) and Long-Low (LL; $14-38$ $\mu$m; $\lambda/\Delta\lambda\sim90$) low-resolution modules.
The observations used a ramp time of 6s with 5 cycles of standard
pointed observations for V348 Sgr and 10 cycles for HV 2671. So each of the bands (SL1, SL2, LL1 and LL2) was integrated for 63s for V348 Sgr and 126s for HV 2671.
V348 Sgr was observed on 2006 October 22, and  HV 2671 was observed on 2006 November 14. 

Our point-source extraction and calibration method has been extensively employed in  \citet{Furlan:2006fj}, \citet{Sargent:2006kx}, \citet{Watson:2007qy}, and \citet{Sargent:2009uq} to which we refer the reader for more detail.  We start with the SSC IRS pipeline {\it Basic Calibrated Data} (BCD) product for each object.  The BCD is flat-fielded, dark-current subtracted, and stray-light corrected.  We employ a point-source spectral extraction and calibration method using the Spectral Modeling, Analysis and Reduction Tool (SMART; \citealt{Higdon:2004lr}) and IDL routines for post-pipeline processing.  In particular, we identify and correct for rogue (NaN) pixels in our two dimensional spectral data by linearly interpolating the four nearest neighboring pixel values.  To correct for the sky background, we subtract the off-order spectrum ($\sim$1-3\arcmin~away from target) in the same nod position of the on-target order.  The low-resolution sky-subtracted spectra are then extracted using a variable-width extraction window that fits tightly to the IRS point-spread function.

To calibrate our spectra, we employ custom relative spectral response functions (RSRFs) which yield flux densities based on the signal detected at a given wavelength.  To produce the RSRFs, we use SMART to divide each of our spectra nod by nod for each order of each module.  A spectral template of a calibration star ($\alpha$ Lac; \citealt{Cohen:2003fk}) was identically prepared such that the quotient of the template and the observed stellar spectrum is the RSRF.  The low-resolution, sky-subtracted extractions were then multiplied by the RSRF corresponding to the relevant nod, order and module.  In general, we found good agreement between flux values in regions of wavelength overlap between orders in each module.  For each object, the procedure described above produces a calibrated point-source spectrum.
For V348 Sgr, there was a problem with the SL2 nod 1 data which caused a mismatch in the 8 \micron~region.  So only the SL2 nod 2 data were used.  
The small apparent emission features seen in the  21 \micron~region for V348 Sgr are not real but are the result of splicing the spectra. 

Figure 1 shows the fully reduced IRS spectra of V348 Sgr and HV 2671. The ISO spectrum of CPD --56$^{\rm o}$~8032 is also plotted for comparison \citep{1999ApJ...513L.135C}. 
To enhance the emission features, the underlying continua have been fitted and removed. It was not possible to fit any of the three continua with a single modified blackbody. The dust around these stars has a distribution of temperatures so a power law was used to fit the continuum of each star which was then subtracted \citep{2011ApJ...733...93G}.
The emission shown in Figure 1 comes from a region of $\sim$3$\farcs$6 and 10$\farcs$2 for the SL and LL modules, respectively.
The ISO SWS apertures are quite large ($\geq$14\arcsec x 20\arcsec).
Two ISO spectra exist for V348 Sgr but they have very low S/N \citep{2003ApJS..147..379S}. They will not be considered further here. 
There is no sign of extended emission in the IRS spectra of HV 2671 and V348 Sgr.

In Figures 2-4, the SEDs of HV 2671, V348 Sgr and  CPD --56$^{\rm o}$~8032 are plotted from the visible to the far-IR. The data, summarized in Tables 1-3,  consist of visible photometry, 2MASS JHK, as well as IRAS photometry  \citep{1985A&A...152...58W, 1986MNRAS.222..357K}. 
AKARI/IRC and FIS photometry are also available for the three stars \citep{2007PASJ...59S.369M,2010A&A...514A...1I}.
For HV 2671, Spitzer and Herschel photometry are also available. 
The Spitzer/IRAC and MIPS 24 \micron~photometry were obtained from the catalogs provided by the SAGE Legacy project \citep{2006AJ....132.2268M}. 
MIPS spectra (52-93 \micron) were also obtained for HV 2671 \citep{2010AJ....139...68V}.
Herschel data for HV 2671 were obtained as part of the HERITAGE Key Program \citep{2010A&A...518L..71M}.
Fluxes were extracted from  PACS (100, 160 \micron) and SPIRE (250 \micron) using Starfinder \citep{2000A&AS..147..335D}. 
Starfinder produced point spread function (PSF)  fitted 
photometry using PSFs for each instrument which were downloaded from the Herschel Science
Center.  No aperture corrections were applied since the PSFs used were of sufficiently
large diameter.
The uncertainties quoted in Table 1 and plotted in Figure 2 for PACS and SPIRE are the Starfinder uncertainties plus calibration errors of 5\% for SPIRE and 10\% for PACS.  
Figure 5 shows the HV 2671 source in the Spitzer/IRAC 8.0 \micron, Spitzer/MIPS 24 \micron, and SPIRE 250 \micron~bands.
HV 2671 is also visible on the SPIRE 350 and 500 \micron~images but photometry was not attempted because the field is confused with diffuse emission.

\section{Dust Emission Features in the Mid-IR}


The optical spectra of CPD --56$^{\rm o}$~8032, V348 Sgr, and HV 2671 can be found in \citet{1998MNRAS.296..419D}, \citet{1994A&AS..103..445L}, and \citet{2002AJ....123.3387D}, respectively. The stellar spectra exhibit narrow emission lines. In the case of CPD --56$^{\rm o}$~8032 the spectrum is dominated by C IV and C III, while in the case of V348 Sgr, C IV is absent and the spectrum is dominated by neutral and singly ionized species, denoting a cooler atmosphere \citep{1994A&A...283..567L}. The abundances of these stars, determined by non-LTE, spherical wind models are very similar indeed. Based solely on the abundances and the appearance of the stellar spectrum one would deduce that the cooler stars have left the AGB more recently and, given time, they will heat up to look more like CPD --56$^{\rm o}$~8032. 

The IR spectra, however, seem to tell a different story. In Figure 1, CPD --56$^{\rm o}$~8032 clearly shows the usual PAH emission bands at 6.2, 7.7, 8.7 and 11.3 \micron~along with the emission ``plateaus" at 6--9 and 11--14 \micron~\citep{1999ApJ...513L.135C}. The ISO spectrum also shows the 3.3 and 5.2 \micron~PAH features, not plotted here. The spectrum of HV 2671 mimics the features of CPD --56$^{\rm o}$~8032 very closely so that from 5 to 15 \micron~the spectra are close to identical. Also identified in Figure 1, are the crystalline silicate features of Enstatite and Forsterite in the CPD --56$^{\rm o}$~8032 spectrum \citep{2002MNRAS.332..879C}. HV 2671 shows no significant emission from crystalline silicates although it may show the same unidentified emission feature at 32.8 \micron. 
 
The spectrum of V348 Sgr is very different from those of CPD --56$^{\rm o}$~8032 and HV 2671. 
A continuum consisting of a 725 K blackbody and a 63 K modified blackbody was subtracted to produce the spectrum of V348 Sgr shown in Figure 1 \citep{2001ApJ...555..925L,2011ApJ...733...93G}.
There is no PAH emission. There is a weak emission feature present at 6.4 \micron. The spectrum is dominated by a broad emission centered at about 8.2 \micron. This feature is not identified with either PAHs or silicates. 
The V348 Sgr spectrum is quite similar to the average spectrum of nine RCB stars \citep{2011ApJ...729..126G}. The broad feature at $\sim$8 \micron~has been attributed to amorphous carbon \citep{1995A&AS..113..561C} and  hydrogenated amorphous carbon \citep{2001ApJ...555..925L}.
Nova DZ Cru shows a similar spectrum to V348 Sgr with a broad feature at 8.2 \micron~and no other strong features~\citep{2010MNRAS.406L..85E}. The feature is attributed to hydrogenated amorphous carbon (AC) with aliphatic bands. Several other novae and post-AGB stars have shown similar features in their IR spectra \citep{2010MNRAS.406L..85E}. 
The post-AGB stars fall into the Class {\it C} which has  broad emission features at $\sim$6.3 and 8.2  \micron~ascribed to C-C modes \citep{2007ApJ...664.1144S}.  
The luminous transient, NGC 300-OT, also shows a similar mid-IR spectrum with a strong emission feature at 
$\sim$8.5 \micron~\citep{2009ApJ...705.1425P}.
Both HV 2671 and V348 Sgr have two atomic emission lines from ionized elements, likely deriving from the PN ([Si II] at $\sim$34.8 \micron~and [S III] at $\sim$33.4 \micron).  This is evidence of a PN round HV 2671. There also is a possible Fe II line at 35.8  \micron~in both objects. 

Recently, emission features associated with C$_{60}$ have been identified in in several PNe  \citep{2010Sci...329.1180C,2010ApJ...724L..39G}, a proto-PN \citep{2011ApJ...730..126Z}, in the interstellar medium \citep{2010ApJ...722L..54S}, and in the least H-deficient RCB stars, DY Cen and V854 Cen \citep{2011ApJ...729..126G}.
DY Cen is a hot RCB star like V348 Sgr and HV 2671. However, as mentioned above, the helium abundances of V348 Sgr and HV 2671 are more similar to [WC] CSPNe like CPD --56$^{\rm o}$~8032 than to the cooler RCB stars, while DY Cen's helium abundance resembles the cooler RCB stars. The  C$_{60}$ features identified in \citet{2011ApJ...729..126G} are marked in Figure 1. The features  appear at 7.0, 8.5, 17.4, and 18.9 \micron. There is also a PAH feature at 8.6 \micron.
The spectrum of CPD --56$^{\rm o}$~8032 shows evidence of C$_{60}$ emission. However, the other two stars show no such emission. 
If these are C$_{60}$ features in CPD --56$^{\rm o}$~8032, then the 17.4 and 18.9 \micron~features are broader than those seen in Tc 1 but still narrower than dust grain features. They are more similar to those seen in the RCB stars, DY Cen and V854 Cen \citep{2010Sci...329.1180C,2011ApJ...729..126G}. Like in V854 Cen,  the 7.0 and 17.4 \micron~features are blended with several PAH emission features so it is not possible to measure the line ratio accurately \citep{2011ApJ...729..126G}.  But comparing the peak emissions in the 7.0/18.9, 8.5/18.9, and 17.4/18.9 ratios gives values that are consistent with the observations and predictions of C$_{60}$ \citep{2010ApJ...722L..54S}. It is interesting that C$_{60}$ may be detected in the least hydrogen-deficient RCB stars since CPD --56$^{\rm o}$~8032, while hydrogen deficient, still has a significant hydrogen abundance \citep{1998MNRAS.296..419D}. \citet{2011ApJ...729..126G} suggest that C$_{60}$ may form from the decomposition of hydrogenated amorphous carbon in moderately hydrogen-deficient circumstellar shells. 

In conclusion, while the dust spectrum of CPD --56$^{\rm o}$~8032, with the dual-chemistry signature, may well be indicating the presence of a disk that formed by binary action in the early AGB phase when the star was oxygen-rich, the other spectra are not obviously explained by this scenario \citep{1999ApJ...513L.135C,2002MNRAS.332..879C}. HV 2671 is similar to CPD --56$^{\rm o}$~8032 in its dust properties, but it has no sign of silicate emission. It is possible that the silicate features are just weaker. V348 Sgr, however, is completely different. We may have expected it to be similar to HV 2671 on the grounds of their almost identical optical spectra, but it is not. Its dust emission is difficult to interpret. The resemblance of V348 Sgr to some novae establishes only one weak and circumstantial connection to this class of binaries, a connection that has been established between the [WC] spectra class and ONeMg novae on the grounds of high neon abundances found in the ejecta of the [WC] V605~Aql and two other related stars \citep{2003MNRAS.340..253W,2008MNRAS.383.1639W,2011MNRAS.410.1870L}, and between post-common envelope CSPN and novae \citep{2010MNRAS.407L..21R}. A more solid link between [WC] stars, RCBs remains elusive and the presence or absence of dual-dust chemistry is not a definitive clue to binarity. 


\section{Radiative Transfer Modeling of the Circumstellar Dust}

A comparison of the IRAS and ISO data from the 80's and 90's with the newer Spitzer and AKARI data 
indicates that the mid-IR emission does not vary significantly over these timescales in any of the three stars.
No significant flux variation is seen between the two IRAC/SAGE epochs for HV 2671 \citep{2006AJ....132.2268M}.
Previous modeling showed that the IRAS photometry for V348 Sgr can be fit with a 550 K black body \citep{1985A&A...152...58W}, but
\citet{1986MNRAS.222..357K} note that the near-IR and IRAS photometry cannot be fit by a single black body. The 60 \micron~point for V348 Sgr sits above the $\sim$600 K black body fit to the other data, and indicates the presence of a cooler dust component. 
\citet{2009A&A...501..985T} suggest that the HV 2671 dust is thick and cold compared to other RCB stars based on Spitzer IRAC and MIPS colors.

The availability of photometry for the three stars from the optical to the far-IR provides an opportunity to model the radiative transfer (RT) in their circumstellar dust shells. We have modeled the SEDs of the three stars using a Monte Carlo RT code which includes nonisotropic scattering, polarization, and thermal emission from dust in a spherical-polar grid \citep{2003ApJ...598.1079W,2003ApJ...591.1049W,2006ApJS..167..256R}. All of the models were done with amorphous carbon dust and an MRN size distribution \citep{1977ApJ...217..425M}. We have assumed a mass gas-to-dust ratio of 100.  The best fits to the SEDs are shown in Figures 2-4.  The best fitting model for all three stars consists of a flared disk surrounded by a large low-density envelope (5 $\times $10$^{-21}$ g cm$^{-3}$  in the outer regions, with an $r^{-2}$ radial dependency) that includes a bipolar, lower-density cavity (1$\times$10$^{-21}$ g cm$^{-3}$). 
Heating by the interstellar radiation field was included but was found to be negligible. The stellar parameters and circumstellar dust parameters are summarized in Table 4, where disk and envelope masses are based on the dust mass and a gas-to-dust ratio of 100. The fits to the SEDs of HV~2671, V348~Sgr and CPD --56$^{\rm o}$~8032 are shown in Figures 2, 3 and 4, respectively.
These models are not unique but do provide a general idea of the dust geometry around these stars.

The SEDs of all three sources are fit with a combination of a circumstellar disk and a more extended envelope partially evacuated in the polar regions. The disk provides mid-IR emission from warm dust near the star, the extended envelope supplies the far-IR emission from cooler dust, and the bipolar cavities allow optical radiation to escape with relatively low extinction.  The envelope surrounding V348 is relatively small and optically thin.  The envelopes surrounding the other two stars are much larger and more opaque, with bipolar cavities subtending 40\degree-- 50\degree~(Table 4).  The large-scale dust may or may not be associated with the stars, as discussed below.
The inner hot dust can be fitted by a disk or a patchy distribution of dust, but a complete shell would increase the extinction more than allowed by the data. A disk is consistent  with the dust geometry inferred for CPD --56$^{\rm o}$~8032 by {\it Hubble Space Telescope} \citep{2002ApJ...574L..83D} as well as with the model of an oxygen rich disk within a carbon rich, more distributed dusty environment \citep{1999ApJ...513L.135C}.  The geometry of the dust around the RCB stars in not certain but there is some evidence for disk-like geometries \citep[e.g.,][]{1993A&A...274..330K,1997ApJ...476..870C,2011MNRAS.tmp..621B}. A patchy distribution of dust is consistent with the production of dust in puffs in the RCB stars \citep{1997MNRAS.285..317F}.

The best fit to the SED of V348 Sgr is not as successful as those for the other two stars. This is mainly due to the presence of the strong broad emission feature at $\sim$8.2 \micron~in the IRS spectrum of V348 Sgr which is not accounted for in the radiative transfer which models only the continuum dust emission due to amorphous carbon grains. 
The SED of V348~Sgr is fit with a disk and envelope like HV 2671 and CPD --56$^{\rm o}$~8032 but much smaller in extent and mass. In this sense, V348~Sgr sets itself apart from the other two objects.  
 
The total envelope mass inferred for HV 2671 and, to a lesser extent for CPD --56$^{\rm o}$~8032, is large and would imply relatively high mass stars  \citep[We recall that the likely median progenitor mass for PN is 1.2~M$_\odot$,][]{2006ApJ...650..916M}. However, not all the mass in the low-density envelope may have come from the star itself. HV~2671 lies in an area of strong diffuse dust emission at far-IR wavelengths and it is a hot, luminous star that will heat up a large volume of low-density ISM. It should be noted that since HV 2671 is in the LMC a resolution element will sample a volume $\geq$ 1000 times greater than for the two Galactic stars. CPD --56$^{\rm o}$~8032 also needs a large diffuse envelope for its RT models. The inferred densities of these envelopes, while low, are still orders of magnitudes higher than the typical diffuse ISM, so this material may have been swept up by the material lost from these stars. Similar shells have been seen around other evolved stars including the final-flash star, V605~Aql \citep{1993ApJ...409..725Y,1997AJ....114.2679C}. R CrB, itself, shows a very large shell seen at IRAS 100 \micron~\citep{1986ApJ...310..842G}. That shell has a total shell mass of $\sim$1 M$_{\sun}$ and a density of 0.07 cm$^{-3}$. It is larger and more diffuse than the shell around HV 2671. 
The model shown in Figure 2 does not fit the SPIRE 250 \micron~point very well. We could improve the fit by placing low-density material in a shell at 5 pc radius from the star.  The mass of such a large shell is enormous, 58,000 M$_{\sun}$.  This suggests that the dust contributing to the 250 \micron~point is not associated with the star.
 
 The HV 2671 dust shows no sign of silicates and is fit well by an RT model with AC dust. The dust around CPD --56$^{\rm o}$~8032, which contains both silicates and AC, is nevertheless fit extremely well by the AC only dust. Models with a mixture of silicates and AC  achieve similar fits to the SED of CPD --56$^{\rm o}$~8032. This shows that the correct geometry is more important to a good fit than the dust properties. 
 
Finally, this is the first time that we are detecting cold dust heated by these stars. Despite the very large declines in brightness ($\sim$8 mag) seen in the RCB stars, the amount of dust responsible for each decline is small. The dust is thought to form in small clumps around the star which if along the line of sight cause a decline. These clumps are $\sim$10$^{-8}$ M$_{\sun}$ and the mass loss rate per year is about ten times larger than that \citep{1992ApJ...397..652C,1996PASP..108..225C}. The lifetime of an RCB star is unknown but may be 10$^4$ to 10$^5$ yr \citep[e.g.,][]{2005AJ....130.2293Z}. So assuming a normal gas-to-dust ratio, an RCB star could lose $\sim$1 M$_{\sun}$ in its lifetime. Therefore, the large difference between the amount of gas and dust around HV 2671 and V348 Sgr may have more to do with the ISM environment of the stars, in particular since the the inner dust properties of these two stars are not dissimilar. 

\section{Summary}

We have observed the IR spectra of two of the four hot RCB stars, V348 Sgr and its apparent twin in the LMC, HV 2671. These two objects may constitute a link between the RCB stars and the  [WC] class of CSPNe that has no hydrogen in their atmospheres. We have also observed the IR spectrum of the [WC] star, CPD --56$^{\rm o}$~8032, which also shows RCB-type declines. We expected that the dust chemistries of these three stars might be similar, and reveal further links between the two classes that would elucidate their possible common ancestry. In particular, detecting the presence of dual-dust chemistry would have established a much stronger link between RCB stars and the [WC] central stars of PNe. 
This has not happened, however. The spectrum of HV 2671 is somewhat similar to that of CPD --56$^{\rm o}$~8032, but it lacks the strong crystalline silicate features that we ascribe to the disk of CPD --56$^{\rm o}$~8032. It could be that these features are simply weaker and that the two dusty environments are actually quite similar. On the other hand, the spectrum of V348 Sgr is completely different from those of the other two stars. This spectrum is similar to that of cool RCB stars and a handful of novae.
Radiative transfer models of the three dusty environments reveal that the mass of ejected dust and gas must be quite large, in particular for HV~2671. If this massive cool envelope is not swept up interstellar gas, then these findings point to the fact that HV~2671 had a mass at the high end of the intermediate mass-spectrum. 



\acknowledgments

We would like to thank the referee for making suggestions that improved the paper. 
This work was supported by Spitzer Space Telescope RSA 1287524 issued by
Caltech/JPL. 
We acknowledge financial support from the
NASA Herschel Science Center, JPL contract Nos. 1381522 and 1381650.
We thank Warren Reid and Quentin Parker for providing access to their LMC H$\alpha$ survey data. 
This publication makes use of data products from the Two Micron All Sky Survey, which is a joint project of the University of Massachusetts and the Infrared Processing and Analysis Center/California Institute of Technology, funded by the National Aeronautics and Space Administration and the National Science Foundation.  
This research is based on observations with AKARI, a JAXA project with the participation of ESA.
We appreciate the contributions
and support from the European Space Agency (ESA), the PACS and SPIRE
teams, the Herschel Science Center and the NASA Herschel Science Center (esp.\
A. Barbar and K. Xu) and the PACS/SPIRE instrument control center at CEA-Saclay,
which made this work possible.


\bibliography{/Users/gclayton/projects/latexstuff/everything2}

\begin{thebibliography}{84}
\expandafter\ifx\csname natexlab\endcsname\relax\def\natexlab#1{#1}\fi

\bibitem[{{Alcock} {et~al.}(1996){Alcock}, {Allsman}, {Alves}, {Axelrod},
  {Becker}, {Bennett}, {Clayton}, {Cook}, {Freeman}, {Griest}, {Guern},
  {Kilkenny}, {Lehner}, {Marshall}, {Minniti}, {Peterson}, {Pratt}, {Quinn},
  {Rodgers}, {Stubbs}, {Sutherland}, \& {Welch}}]{1996ApJ...470..583A}
{Alcock}, C., {et~al.} 1996, \apj, 470, 583

\bibitem[{{Alcock et al.}(2001)}]{2001ApJ...554..298A}
{Alcock et al.} 2001, \apj, 554, 298

\bibitem[{{Asplund} {et~al.}(2000){Asplund}, {Gustafsson}, {Lambert}, \&
  {Rao}}]{2000A&A...353..287A}
{Asplund}, M., {Gustafsson}, B., {Lambert}, D.~L., \& {Rao}, N.~K. 2000, \aap,
  353, 287

\bibitem[{{Barlow}(1997)}]{1997Ap&SS.251...15B}
{Barlow}, M.~J. 1997, \apss, 251, 15

\bibitem[{{Bright} {et~al.}(2011){Bright}, {Chesneau}, {Clayton}, {de Marco},
  {Le{\~a}o}, {Nordhaus}, \& {Gallagher}}]{2011MNRAS.tmp..621B}
{Bright}, S.~N., {Chesneau}, O., {Clayton}, G.~C., {de Marco}, O., {Le{\~a}o},
  I.~C., {Nordhaus}, J., \& {Gallagher}, J.~S. 2011, \mnras, 621

\bibitem[{{Cami} {et~al.}(2010){Cami}, {Bernard-Salas}, {Peeters}, \&
  {Malek}}]{2010Sci...329.1180C}
{Cami}, J., {Bernard-Salas}, J., {Peeters}, E., \& {Malek}, S.~E. 2010,
  Science, 329, 1180

\bibitem[{{Chesneau} {et~al.}(2006){Chesneau}, {Collioud}, {De Marco}, {Wolf},
  {Lagadec}, {Zijlstra}, {Rothkopf}, {Acker}, {Clayton}, \&
  {Lopez}}]{2006A&A...455.1009C}
{Chesneau}, O., {et~al.} 2006, \aap, 455, 1009

\bibitem[{{Clayton}(1996)}]{1996PASP..108..225C}
{Clayton}, G.~C. 1996, \pasp, 108, 225

\bibitem[{{Clayton} {et~al.}(1997){Clayton}, {Bjorkman}, {Nordsieck},
  {Zellner}, \& {Schulte-Ladbeck}}]{1997ApJ...476..870C}
{Clayton}, G.~C., {Bjorkman}, K.~S., {Nordsieck}, K.~H., {Zellner}, N.~E.~B.,
  \& {Schulte-Ladbeck}, R.~E. 1997, \apj, 476, 870

\bibitem[{{Clayton} \& {De Marco}(1997)}]{1997AJ....114.2679C}
{Clayton}, G.~C., \& {De Marco}, O. 1997, \aj, 114, 2679

\bibitem[{{Clayton} {et~al.}(2007){Clayton}, {Geballe}, {Herwig}, {Fryer}, \&
  {Asplund}}]{2007ApJ...662.1220C}
{Clayton}, G.~C., {Geballe}, T.~R., {Herwig}, F., {Fryer}, C., \& {Asplund}, M.
  2007, \apj, 662, 1220

\bibitem[{{Clayton} {et~al.}(2005){Clayton}, {Herwig}, {Geballe}, {Asplund},
  {Tenenbaum}, {Engelbracht}, \& {Gordon}}]{2005ApJ...623L.141C}
{Clayton}, G.~C., {Herwig}, F., {Geballe}, T.~R., {Asplund}, M., {Tenenbaum},
  E.~D., {Engelbracht}, C.~W., \& {Gordon}, K.~D. 2005, \apjl, 623, L141

\bibitem[{{Clayton} {et~al.}(2006){Clayton}, {Kerber}, {Pirzkal}, {De Marco},
  {Crowther}, \& {Fedrow}}]{2006ApJ...646L..69C}
{Clayton}, G.~C., {Kerber}, F., {Pirzkal}, N., {De Marco}, O., {Crowther},
  P.~A., \& {Fedrow}, J.~M. 2006, \apjl, 646, L69

\bibitem[{{Clayton} {et~al.}(1992){Clayton}, {Whitney}, {Stanford}, \&
  {Drilling}}]{1992ApJ...397..652C}
{Clayton}, G.~C., {Whitney}, B.~A., {Stanford}, S.~A., \& {Drilling}, J.~S.
  1992, \apj, 397, 652

\bibitem[{{Cohen} {et~al.}(2002){Cohen}, {Barlow}, {Liu}, \&
  {Jones}}]{2002MNRAS.332..879C}
{Cohen}, M., {Barlow}, M.~J., {Liu}, X.-W., \& {Jones}, A.~F. 2002, \mnras,
  332, 879

\bibitem[{{Cohen} {et~al.}(1999){Cohen}, {Barlow}, {Sylvester}, {Liu}, {Cox},
  {Lim}, {Schmitt}, \& {Speck}}]{1999ApJ...513L.135C}
{Cohen}, M., {Barlow}, M.~J., {Sylvester}, R.~J., {Liu}, X.-W., {Cox}, P.,
  {Lim}, T., {Schmitt}, B., \& {Speck}, A.~K. 1999, \apjl, 513, L135

\bibitem[{{Cohen} {et~al.}(2003){Cohen}, {Megeath}, {Hammersley},
  {Mart{\'{\i}}n-Luis}, \& {Stauffer}}]{Cohen:2003fk}
{Cohen}, M., {Megeath}, S.~T., {Hammersley}, P.~L., {Mart{\'{\i}}n-Luis}, F.,
  \& {Stauffer}, J. 2003, \aj, 125, 2645

\bibitem[{{Cohen} {et~al.}(1989){Cohen}, {Tielens}, {Bregman}, {Witteborn},
  {Rank}, {Allamandola}, {Wooden}, \& {Jourdain de
  Muizon}}]{1989ApJ...341..246C}
{Cohen}, M., {Tielens}, A.~G.~G.~M., {Bregman}, J., {Witteborn}, F.~C., {Rank},
  D.~M., {Allamandola}, L.~J., {Wooden}, D., \& {Jourdain de Muizon}, M. 1989,
  \apj, 341, 246

\bibitem[{{Colangeli} {et~al.}(1995){Colangeli}, {Mennella}, {Palumbo},
  {Rotundi}, \& {Bussoletti}}]{1995A&AS..113..561C}
{Colangeli}, L., {Mennella}, V., {Palumbo}, P., {Rotundi}, A., \& {Bussoletti},
  E. 1995, \aaps, 113, 561

\bibitem[{{Crowther} {et~al.}(1998){Crowther}, {De Marco}, \&
  {Barlow}}]{1998MNRAS.296..367C}
{Crowther}, P.~A., {De Marco}, O., \& {Barlow}, M.~J. 1998, \mnras, 296, 367

\bibitem[{{De Marco}(2008)}]{2008ASPC..391..209D}
{De Marco}, O. 2008, in Astronomical Society of the Pacific Conference Series,
  Vol. 391, Hydrogen-Deficient Stars, ed. {A.~Werner \& T.~Rauch}, 209

\bibitem[{{De Marco} \& {Barlow}(2001)}]{2001Ap&SS.275...53D}
{De Marco}, O., \& {Barlow}, M.~J. 2001, \apss, 275, 53

\bibitem[{{De Marco} {et~al.}(2002{\natexlab{a}}){De Marco}, {Barlow}, \&
  {Cohen}}]{2002ApJ...574L..83D}
{De Marco}, O., {Barlow}, M.~J., \& {Cohen}, M. 2002{\natexlab{a}}, \apjl, 574,
  L83

\bibitem[{{De Marco} {et~al.}(1997){De Marco}, {Barlow}, \&
  {Storey}}]{1997MNRAS.292...86D}
{De Marco}, O., {Barlow}, M.~J., \& {Storey}, P.~J. 1997, \mnras, 292, 86

\bibitem[{{De Marco} {et~al.}(2002{\natexlab{b}}){De Marco}, {Clayton},
  {Herwig}, {Pollacco}, {Clark}, \& {Kilkenny}}]{2002AJ....123.3387D}
{De Marco}, O., {Clayton}, G.~C., {Herwig}, F., {Pollacco}, D.~L., {Clark},
  J.~S., \& {Kilkenny}, D. 2002{\natexlab{b}}, \aj, 123, 3387

\bibitem[{{De Marco} \& {Crowther}(1998)}]{1998MNRAS.296..419D}
{De Marco}, O., \& {Crowther}, P.~A. 1998, \mnras, 296, 419

\bibitem[{{De Marco} \& {Soker}(2002)}]{2002PASP..114..602D}
{De Marco}, O., \& {Soker}, N. 2002, \pasp, 114, 602

\bibitem[{{Diolaiti} {et~al.}(2000){Diolaiti}, {Bendinelli}, {Bonaccini},
  {Close}, {Currie}, \& {Parmeggiani}}]{2000A&AS..147..335D}
{Diolaiti}, E., {Bendinelli}, O., {Bonaccini}, D., {Close}, L., {Currie}, D.,
  \& {Parmeggiani}, G. 2000, \aaps, 147, 335

\bibitem[{{Evans} {et~al.}(2010){Evans}, {Gehrz}, {Woodward}, {Helton},
  {Rushton}, {Bode}, {Krautter}, {Lyke}, {Lynch}, {Ness}, {Starrfield},
  {Truran}, \& {Wagner}}]{2010MNRAS.406L..85E}
{Evans}, A., {et~al.} 2010, \mnras, 406, L85

\bibitem[{{Feast} {et~al.}(1997){Feast}, {Carter}, {Roberts}, {Marang}, \&
  {Catchpole}}]{1997MNRAS.285..317F}
{Feast}, M.~W., {Carter}, B.~S., {Roberts}, G., {Marang}, F., \& {Catchpole},
  R.~M. 1997, \mnras, 285, 317

\bibitem[{{Furlan} {et~al.}(2006){Furlan}, {Hartmann}, {Calvet}, {D'Alessio},
  {Franco-Hern{\'a}ndez}, {Forrest}, {Watson}, {Uchida}, {Sargent}, {Green},
  {Keller}, \& {Herter}}]{Furlan:2006fj}
{Furlan}, E., {et~al.} 2006, \apjs, 165, 568

\bibitem[{{Garc{\'{\i}}a-Hern{\'a}ndez}
  {et~al.}(2009){Garc{\'{\i}}a-Hern{\'a}ndez}, {Hinkle}, {Lambert}, \&
  {Eriksson}}]{2009ApJ...696.1733G}
{Garc{\'{\i}}a-Hern{\'a}ndez}, D.~A., {Hinkle}, K.~H., {Lambert}, D.~L., \&
  {Eriksson}, K. 2009, \apj, 696, 1733

\bibitem[{{Garc{\'{\i}}a-Hern{\'a}ndez}
  {et~al.}(2011){Garc{\'{\i}}a-Hern{\'a}ndez}, {Kameswara Rao}, \&
  {Lambert}}]{2011ApJ...729..126G}
{Garc{\'{\i}}a-Hern{\'a}ndez}, D.~A., {Kameswara Rao}, N., \& {Lambert}, D.~L.
  2011, \apj, 729, 126

\bibitem[{{Garc{\'{\i}}a-Hern{\'a}ndez}
  {et~al.}(2010{\natexlab{a}}){Garc{\'{\i}}a-Hern{\'a}ndez}, {Lambert},
  {Kameswara Rao}, {Hinkle}, \& {Eriksson}}]{2010ApJ...714..144G}
{Garc{\'{\i}}a-Hern{\'a}ndez}, D.~A., {Lambert}, D.~L., {Kameswara Rao}, N.,
  {Hinkle}, K.~H., \& {Eriksson}, K. 2010{\natexlab{a}}, \apj, 714, 144

\bibitem[{{Garc{\'{\i}}a-Hern{\'a}ndez}
  {et~al.}(2010{\natexlab{b}}){Garc{\'{\i}}a-Hern{\'a}ndez}, {Manchado},
  {Garc{\'{\i}}a-Lario}, {Stanghellini}, {Villaver}, {Shaw}, {Szczerba}, \&
  {Perea-Calder{\'o}n}}]{2010ApJ...724L..39G}
{Garc{\'{\i}}a-Hern{\'a}ndez}, D.~A., {Manchado}, A., {Garc{\'{\i}}a-Lario},
  P., {Stanghellini}, L., {Villaver}, E., {Shaw}, R.~A., {Szczerba}, R., \&
  {Perea-Calder{\'o}n}, J.~V. 2010{\natexlab{b}}, \apjl, 724, L39

\bibitem[{{Gillett} {et~al.}(1986){Gillett}, {Backman}, {Beichman}, \&
  {Neugebauer}}]{1986ApJ...310..842G}
{Gillett}, F.~C., {Backman}, D.~E., {Beichman}, C., \& {Neugebauer}, G. 1986,
  \apj, 310, 842

\bibitem[{{Guha Niyogi} {et~al.}(2011){Guha Niyogi}, {Speck}, \&
  {Onaka}}]{2011ApJ...733...93G}
{Guha Niyogi}, S., {Speck}, A.~K., \& {Onaka}, T. 2011, \apj, 733, 93

\bibitem[{{Guzman-Ramirez} {et~al.}(2011){Guzman-Ramirez}, {Zijlstra},
  {N{\'{\i}}chuim{\'{\i}}n}, {Gesicki}, {Lagadec}, {Millar}, \&
  {Woods}}]{2011MNRAS.tmp..441G}
{Guzman-Ramirez}, L., {Zijlstra}, A.~A., {N{\'{\i}}chuim{\'{\i}}n}, R.,
  {Gesicki}, K., {Lagadec}, E., {Millar}, T.~J., \& {Woods}, P.~M. 2011,
  \mnras, 441

\bibitem[{{Heck} {et~al.}(1985){Heck}, {Houziaux}, {Manfroid}, {Jones}, \&
  {Andrews}}]{1985AAS...61..375H}
{Heck}, A., {Houziaux}, L., {Manfroid}, J., {Jones}, D.~H.~P., \& {Andrews},
  P.~J. 1985, \aaps, 61, 375

\bibitem[{{Herwig}(2000)}]{2000A&A...360..952H}
{Herwig}, F. 2000, \aap, 360, 952

\bibitem[{{Herwig}(2001)}]{2001ApJ...554L..71H}
---. 2001, \apjl, 554, L71

\bibitem[{{Higdon} {et~al.}(2004){Higdon}, {Devost}, {Higdon}, {Brandl},
  {Houck}, {Hall}, {Barry}, {Charmandaris}, {Smith}, {Sloan}, \&
  {Green}}]{Higdon:2004lr}
{Higdon}, S.~J.~U., {et~al.} 2004, \pasp, 116, 975

\bibitem[{{Ishihara} {et~al.}(2010){Ishihara}, {Onaka}, {Kataza}, {Salama},
  {Alfageme}, {Cassatella}, {Cox}, {Garc{\'{\i}}a-Lario}, {Stephenson},
  {Cohen}, {Fujishiro}, {Fujiwara}, {Hasegawa}, {Ita}, {Kim}, {Matsuhara},
  {Murakami}, {M{\"u}ller}, {Nakagawa}, {Ohyama}, {Oyabu}, {Pyo}, {Sakon},
  {Shibai}, {Takita}, {Tanab{\'e}}, {Uemizu}, {Ueno}, {Usui}, {Wada},
  {Watarai}, {Yamamura}, \& {Yamauchi}}]{2010A&A...514A...1I}
{Ishihara}, D., {et~al.} 2010, \aap, 514, A1

\bibitem[{{Jones} {et~al.}(1999){Jones}, {Lawson}, {De Marco}, {Kilkenny}, {van
  Wyk}, \& {Roberts}}]{1999Obs...119...76J}
{Jones}, A., {Lawson}, W., {De Marco}, O., {Kilkenny}, D., {van Wyk}, F., \&
  {Roberts}, G. 1999, The Observatory, 119, 76

\bibitem[{{Lambert} \& {Rao}(1994)}]{1994JApA...15...47L}
{Lambert}, D.~L., \& {Rao}, N.~K. 1994, Journal of Astrophysics and Astronomy,
  15, 47

\bibitem[{{Lambert} {et~al.}(2001){Lambert}, {Rao}, {Pandey}, \&
  {Ivans}}]{2001ApJ...555..925L}
{Lambert}, D.~L., {Rao}, N.~K., {Pandey}, G., \& {Ivans}, I.~I. 2001, \apj,
  555, 925

\bibitem[{{Lau} {et~al.}(2011){Lau}, {de Marco}, \&
  {Liu}}]{2011MNRAS.410.1870L}
{Lau}, H.~H.~B., {de Marco}, O., \& {Liu}, X. 2011, \mnras, 410, 1870

\bibitem[{{Leuenhagen} \& {Hamann}(1994)}]{1994A&A...283..567L}
{Leuenhagen}, U., \& {Hamann}, W. 1994, \aap, 283, 567

\bibitem[{{Leuenhagen} {et~al.}(1994){Leuenhagen}, {Heber}, \&
  {Jeffery}}]{1994A&AS..103..445L}
{Leuenhagen}, U., {Heber}, U., \& {Jeffery}, C.~S. 1994, \aaps, 103, 445

\bibitem[{{Mathis} {et~al.}(1977){Mathis}, {Rumpl}, \&
  {Nordsieck}}]{1977ApJ...217..425M}
{Mathis}, J.~S., {Rumpl}, W., \& {Nordsieck}, K.~H. 1977, \apj, 217, 425

\bibitem[{{Meixner} {et~al.}(2006){Meixner}, {Gordon}, {Indebetouw}, {Hora},
  {Whitney}, {Blum}, {Reach}, {Bernard}, {Meade}, {Babler}, {Engelbracht},
  {For}, {Misselt}, {Vijh}, {Leitherer}, {Cohen}, {Churchwell}, {Boulanger},
  {Frogel}, {Fukui}, {Gallagher}, {Gorjian}, {Harris}, {Kelly}, {Kawamura},
  {Kim}, {Latter}, {Madden}, {Markwick-Kemper}, {Mizuno}, {Mizuno}, {Mould},
  {Nota}, {Oey}, {Olsen}, {Onishi}, {Paladini}, {Panagia}, {Perez-Gonzalez},
  {Shibai}, {Sato}, {Smith}, {Staveley-Smith}, {Tielens}, {Ueta}, {van Dyk},
  {Volk}, {Werner}, \& {Zaritsky}}]{2006AJ....132.2268M}
{Meixner}, M., {et~al.} 2006, \aj, 132, 2268

\bibitem[{{Meixner} {et~al.}(2010){Meixner}, {Galliano}, {Hony}, {Roman-Duval},
  {Robitaille}, {Panuzzo}, {Sauvage}, {Gordon}, {Engelbracht}, {Misselt},
  {Okumura}, {Beck}, {Bernard}, {Bolatto}, {Bot}, {Boyer}, {Bracker},
  {Carlson}, {Clayton}, {Chen}, {Churchwell}, {Fukui}, {Galametz}, {Hora},
  {Hughes}, {Indebetouw}, {Israel}, {Kawamura}, {Kemper}, {Kim}, {Kwon},
  {Lawton}, {Li}, {Long}, {Marengo}, {Madden}, {Matsuura}, {Oliveira},
  {Onishi}, {Otsuka}, {Paradis}, {Poglitsch}, {Riebel}, {Reach}, {Rubio},
  {Sargent}, {Sewi{\l}o}, {Simon}, {Skibba}, {Smith}, {Srinivasan}, {Tielens},
  {van Loon}, {Whitney}, \& {Woods}}]{2010A&A...518L..71M}
---. 2010, \aap, 518, L71

\bibitem[{{Moe} \& {De Marco}(2006)}]{2006ApJ...650..916M}
{Moe}, M., \& {De Marco}, O. 2006, \apj, 650, 916

\bibitem[{{Murakami} {et~al.}(2007){Murakami}, {Baba}, {Barthel}, {Clements},
  {Cohen}, {Doi}, {Enya}, {Figueredo}, {Fujishiro}, {Fujiwara}, {Fujiwara},
  {Garcia-Lario}, {Goto}, {Hasegawa}, {Hibi}, {Hirao}, {Hiromoto}, {Hong},
  {Imai}, {Ishigaki}, {Ishiguro}, {Ishihara}, {Ita}, {Jeong}, {Jeong},
  {Kaneda}, {Kataza}, {Kawada}, {Kawai}, {Kawamura}, {Kessler}, {Kester},
  {Kii}, {Kim}, {Kim}, {Kobayashi}, {Koo}, {Kwon}, {Lee}, {Lorente}, {Makiuti},
  {Matsuhara}, {Matsumoto}, {Matsuo}, {Matsuura}, {M{\"u}ller}, {Murakami},
  {Nagata}, {Nakagawa}, {Naoi}, {Narita}, {Noda}, {Oh}, {Ohnishi}, {Ohyama},
  {Okada}, {Okuda}, {Oliver}, {Onaka}, {Ootsubo}, {Oyabu}, {Pak}, {Park},
  {Pearson}, {Rowan-Robinson}, {Saito}, {Sakon}, {Salama}, {Sato}, {Savage},
  {Serjeant}, {Shibai}, {Shirahata}, {Sohn}, {Suzuki}, {Takagi}, {Takahashi},
  {Tanab{\'e}}, {Takeuchi}, {Takita}, {Thomson}, {Uemizu}, {Ueno}, {Usui},
  {Verdugo}, {Wada}, {Wang}, {Watabe}, {Watarai}, {White}, {Yamamura},
  {Yamauchi}, \& {Yasuda}}]{2007PASJ...59S.369M}
{Murakami}, H., {et~al.} 2007, \pasj, 59, 369

\bibitem[{{Perea-Calder{\'o}n} {et~al.}(2009){Perea-Calder{\'o}n},
  {Garc{\'{\i}}a-Hern{\'a}ndez}, {Garc{\'{\i}}a-Lario}, {Szczerba}, \&
  {Bobrowsky}}]{2009A&A...495L...5P}
{Perea-Calder{\'o}n}, J.~V., {Garc{\'{\i}}a-Hern{\'a}ndez}, D.~A.,
  {Garc{\'{\i}}a-Lario}, P., {Szczerba}, R., \& {Bobrowsky}, M. 2009, \aap,
  495, L5

\bibitem[{{Pollacco} {et~al.}(1991){Pollacco}, {Hill}, {Houziaux}, \&
  {Manfroid}}]{1991MNRAS.248P...1P}
{Pollacco}, D.~L., {Hill}, P.~W., {Houziaux}, L., \& {Manfroid}, J. 1991,
  \mnras, 248, 1P

\bibitem[{{Prieto} {et~al.}(2009){Prieto}, {Sellgren}, {Thompson}, \&
  {Kochanek}}]{2009ApJ...705.1425P}
{Prieto}, J.~L., {Sellgren}, K., {Thompson}, T.~A., \& {Kochanek}, C.~S. 2009,
  \apj, 705, 1425

\bibitem[{{Rao} \& {Nandy}(1986)}]{1986MNRAS.222..357K}
{Rao}, N.~K., \& {Nandy}, K. 1986, \mnras, 222, 357

\bibitem[{{Rao} \& {Raveendran}(1993)}]{1993A&A...274..330K}
{Rao}, N.~K., \& {Raveendran}, A.~V. 1993, \aap, 274, 330

\bibitem[{{Reid} \& {Parker}(2006{\natexlab{a}})}]{2006MNRAS.365..401R}
{Reid}, W.~A., \& {Parker}, Q.~A. 2006{\natexlab{a}}, \mnras, 365, 401

\bibitem[{{Reid} \& {Parker}(2006{\natexlab{b}})}]{2006MNRAS.373..521R}
---. 2006{\natexlab{b}}, \mnras, 373, 521

\bibitem[{{Robitaille} {et~al.}(2006){Robitaille}, {Whitney}, {Indebetouw},
  {Wood}, \& {Denzmore}}]{2006ApJS..167..256R}
{Robitaille}, T.~P., {Whitney}, B.~A., {Indebetouw}, R., {Wood}, K., \&
  {Denzmore}, P. 2006, \apjs, 167, 256

\bibitem[{{Rodr{\'{\i}}guez-Gil} {et~al.}(2010){Rodr{\'{\i}}guez-Gil},
  {Santander-Garc{\'{\i}}a}, {Knigge}, {Corradi}, {G{\"a}nsicke}, {Barlow},
  {Drake}, {Drew}, {Miszalski}, {Napiwotzki}, {Steeghs}, {Wesson}, {Zijlstra},
  {Jones}, {Liimets}, {Mu{\~n}oz-Darias}, {Pyrzas}, \&
  {Rubio-D{\'{\i}}ez}}]{2010MNRAS.407L..21R}
{Rodr{\'{\i}}guez-Gil}, P., {et~al.} 2010, \mnras, 407, L21

\bibitem[{{Sabogal} {et~al.}(2005){Sabogal}, {Mennickent}, {Pietrzy{\'n}ski},
  \& {Gieren}}]{2005MNRAS.361.1055S}
{Sabogal}, B.~E., {Mennickent}, R.~E., {Pietrzy{\'n}ski}, G., \& {Gieren}, W.
  2005, \mnras, 361, 1055

\bibitem[{{Sargent} {et~al.}(2006){Sargent}, {Forrest}, {D'Alessio}, {Li},
  {Najita}, {Watson}, {Calvet}, {Furlan}, {Green}, {Kim}, {Sloan}, {Chen},
  {Hartmann}, \& {Houck}}]{Sargent:2006kx}
{Sargent}, B., {et~al.} 2006, \apj, 645, 395

\bibitem[{{Sargent} {et~al.}(2009){Sargent}, {Forrest}, {Tayrien}, {McClure},
  {Li}, {Basu}, {Manoj}, {Watson}, {Bohac}, {Furlan}, {Kim}, {Green}, \&
  {Sloan}}]{Sargent:2009uq}
{Sargent}, B.~A., {et~al.} 2009, \apj, 690, 1193

\bibitem[{{Sellgren} {et~al.}(2010){Sellgren}, {Werner}, {Ingalls}, {Smith},
  {Carleton}, \& {Joblin}}]{2010ApJ...722L..54S}
{Sellgren}, K., {Werner}, M.~W., {Ingalls}, J.~G., {Smith}, J.~D.~T.,
  {Carleton}, T.~M., \& {Joblin}, C. 2010, \apjl, 722, L54

\bibitem[{{Sloan} {et~al.}(2003){Sloan}, {Kraemer}, {Price}, \&
  {Shipman}}]{2003ApJS..147..379S}
{Sloan}, G.~C., {Kraemer}, K.~E., {Price}, S.~D., \& {Shipman}, R.~F. 2003,
  \apjs, 147, 379

\bibitem[{{Sloan} {et~al.}(2007){Sloan}, {Jura}, {Duley}, {Kraemer},
  {Bernard-Salas}, {Forrest}, {Sargent}, {Li}, {Barry}, {Bohac}, {Watson}, \&
  {Houck}}]{2007ApJ...664.1144S}
{Sloan}, G.~C., {et~al.} 2007, \apj, 664, 1144

\bibitem[{{Soszy{\'n}ski} {et~al.}(2009){Soszy{\'n}ski}, {Udalski},
  {Szyma{\'n}ski}, {Kubiak}, {Pietrzy{\'n}ski}, {Wyrzykowski}, {Szewczyk},
  {Ulaczyk}, \& {Poleski}}]{2009AcA....59..335S}
{Soszy{\'n}ski}, I., {et~al.} 2009, Acta Astron., 59, 335

\bibitem[{{Tisserand} {et~al.}(2009){Tisserand}, {Wood}, {Marquette}, {Afonso},
  {Albert}, {Andersen}, {Ansari}, {Aubourg}, {Bareyre}, {Beaulieu}, {Charlot},
  {Coutures}, {Ferlet}, {Fouqu{\'e}}, {Glicenstein}, {Goldman}, {Gould},
  {Gros}, {de Kat}, {Lesquoy}, {Loup}, {Magneville}, {Maurice}, {Maury},
  {Milsztajn}, {Moniez}, {Palanque-Delabrouille}, {Perdereau}, {Rich},
  {Schwemling}, {Spiro}, \& {Vidal-Madjar}}]{2009A&A...501..985T}
{Tisserand}, P., {et~al.} 2009, \aap, 501, 985

\bibitem[{{van Loon} {et~al.}(2010){van Loon}, {Oliveira}, {Gordon}, {Meixner},
  {Shiao}, {Boyer}, {Kemper}, {Woods}, {Tielens}, {Marengo}, {Indebetouw},
  {Sloan}, \& {Chen}}]{2010AJ....139...68V}
{van Loon}, J.~T., {et~al.} 2010, \aj, 139, 68

\bibitem[{{Walker}(1985)}]{1985A&A...152...58W}
{Walker}, H.~J. 1985, \aap, 152, 58

\bibitem[{{Waters} {et~al.}(1998){Waters}, {Beintema}, {Zijlstra}, {de Koter},
  {Molster}, {Bouwman}, {de Jong}, {Pottasch}, \& {de
  Graauw}}]{1998A&A...331L..61W}
{Waters}, L.~B.~F.~M., {et~al.} 1998, \aap, 331, L61

\bibitem[{{Watson} {et~al.}(2007){Watson}, {Bohac}, {Hull}, {Forrest},
  {Furlan}, {Najita}, {Calvet}, {D'Alessio}, {Hartmann}, {Sargent}, {Green},
  {Kim}, \& {Houck}}]{Watson:2007qy}
{Watson}, D.~M., {et~al.} 2007, \nat, 448, 1026

\bibitem[{{Webbink}(1984)}]{1984ApJ...277..355W}
{Webbink}, R.~F. 1984, \apj, 277, 355

\bibitem[{{Wesson} {et~al.}(2008){Wesson}, {Barlow}, {Liu}, {Storey},
  {Ercolano}, \& {de Marco}}]{2008MNRAS.383.1639W}
{Wesson}, R., {Barlow}, M.~J., {Liu}, X., {Storey}, P.~J., {Ercolano}, B., \&
  {de Marco}, O. 2008, \mnras, 383, 1639

\bibitem[{{Wesson} {et~al.}(2003){Wesson}, {Liu}, \&
  {Barlow}}]{2003MNRAS.340..253W}
{Wesson}, R., {Liu}, X., \& {Barlow}, M.~J. 2003, \mnras, 340, 253

\bibitem[{{Whitney} {et~al.}(2003{\natexlab{a}}){Whitney}, {Wood}, {Bjorkman},
  \& {Cohen}}]{2003ApJ...598.1079W}
{Whitney}, B.~A., {Wood}, K., {Bjorkman}, J.~E., \& {Cohen}, M.
  2003{\natexlab{a}}, \apj, 598, 1079

\bibitem[{{Whitney} {et~al.}(2003{\natexlab{b}}){Whitney}, {Wood}, {Bjorkman},
  \& {Wolff}}]{2003ApJ...591.1049W}
{Whitney}, B.~A., {Wood}, K., {Bjorkman}, J.~E., \& {Wolff}, M.~J.
  2003{\natexlab{b}}, \apj, 591, 1049

\bibitem[{{Young} {et~al.}(1993){Young}, {Phillips}, \&
  {Knapp}}]{1993ApJ...409..725Y}
{Young}, K., {Phillips}, T.~G., \& {Knapp}, G.~R. 1993, \apj, 409, 725

\bibitem[{{Zaniewski} {et~al.}(2005){Zaniewski}, {Clayton}, {Welch}, {Gordon},
  {Minniti}, \& {Cook}}]{2005AJ....130.2293Z}
{Zaniewski}, A., {Clayton}, G.~C., {Welch}, D.~L., {Gordon}, K.~D., {Minniti},
  D., \& {Cook}, K.~H. 2005, \aj, 130, 2293

\bibitem[{{Zhang} \& {Kwok}(2011)}]{2011ApJ...730..126Z}
{Zhang}, Y., \& {Kwok}, S. 2011, \apj, 730, 126

\bibitem[{{Zijlstra}(2001)}]{2001ASSL..265..157Z}
{Zijlstra}, A.~A. 2001, in Astrophysics and Space Science Library, Vol. 265,
  Astrophysics and Space Science Library, ed. {R.~Szczerba \& S.~K.~G{\'o}rny},
  157

\end{thebibliography}

\begin{figure}
\figurenum{1} 
\begin{center}
\includegraphics[width=4in,angle=0]{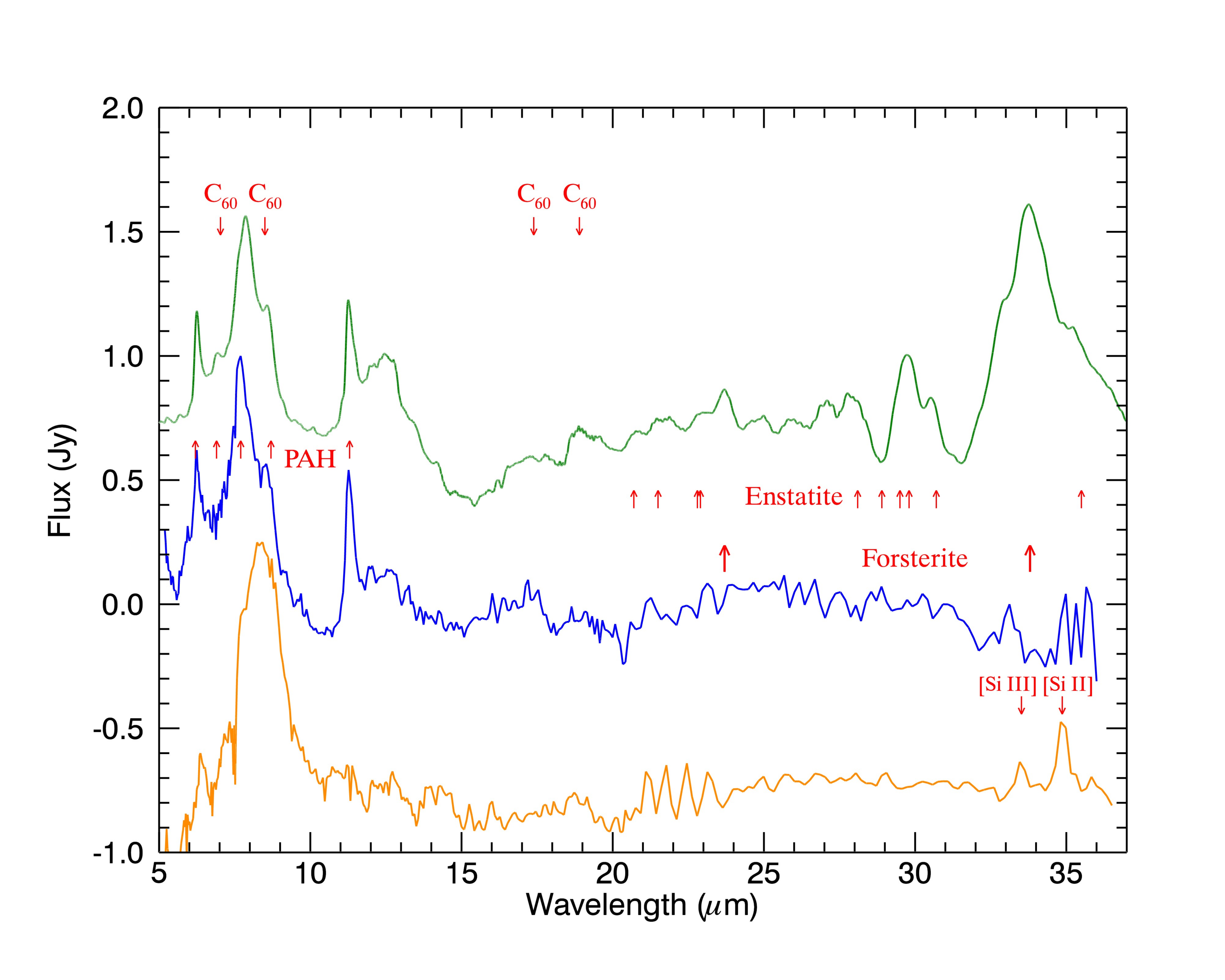}
\end{center}
\caption{Spitzer/IRS spectra of HV 2671 (blue) and V348 Sgr (orange), and the ISO spectrum of CPD --56$^{\rm o}$~8032 (green). The spectra have been continuum subtracted. CPD --56$^{\rm o}$~8032 shows emission features attributable to PAHs, crystalline silicates and C$_{60}$. HV 2671 shows only PAH features. V348 Sgr shows only an unusual feature at $\sim$8.2 \micron.}
\end{figure}

\begin{figure}
\figurenum{2} 
\begin{center}
\includegraphics[width=4in,angle=0]{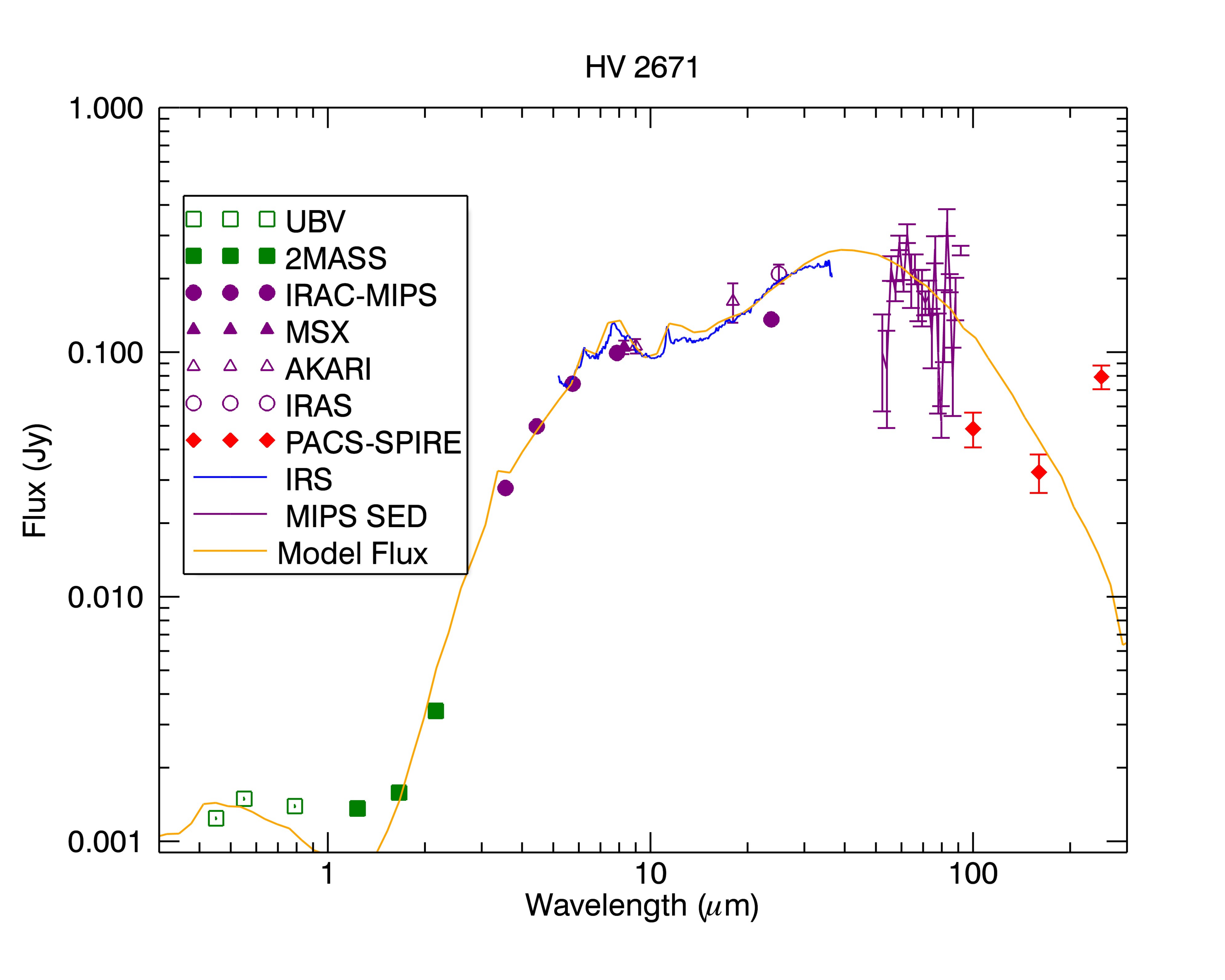}
\end{center}
\caption{SED for HV 2671, and best-fit Monte Carlo RT model. This model consists of amorphous carbon dust in a disk plus a large diffuse envelope. The best fit parameters are shown in Table 4. }
\end{figure}

\begin{figure}
\figurenum{3} 
\begin{center}
\includegraphics[width=4in,angle=0]{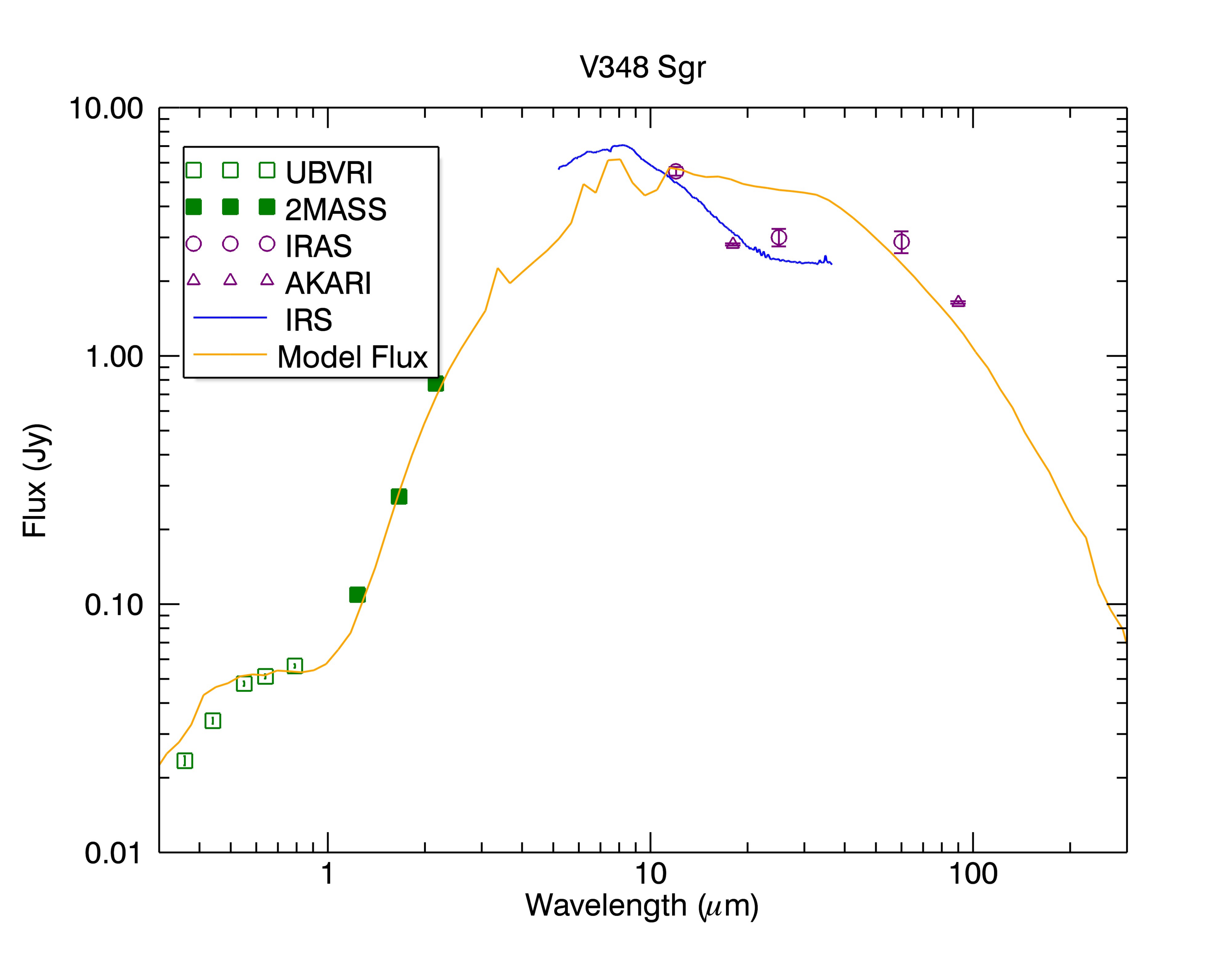}
\end{center}
\caption{SED for V348 Sgr, and best-fit Monte Carlo RT model. This model consists of amorphous carbon dust in a disk plus a large diffuse envelope. The best fit parameters are shown in Table 4. The mismatch seen between the data and the model in the mid-IR may result from the strong emission feature at $\sim$8.2 \micron. }
\end{figure}

\begin{figure}
\figurenum{4} 
\begin{center}
\includegraphics[width=4in,angle=0]{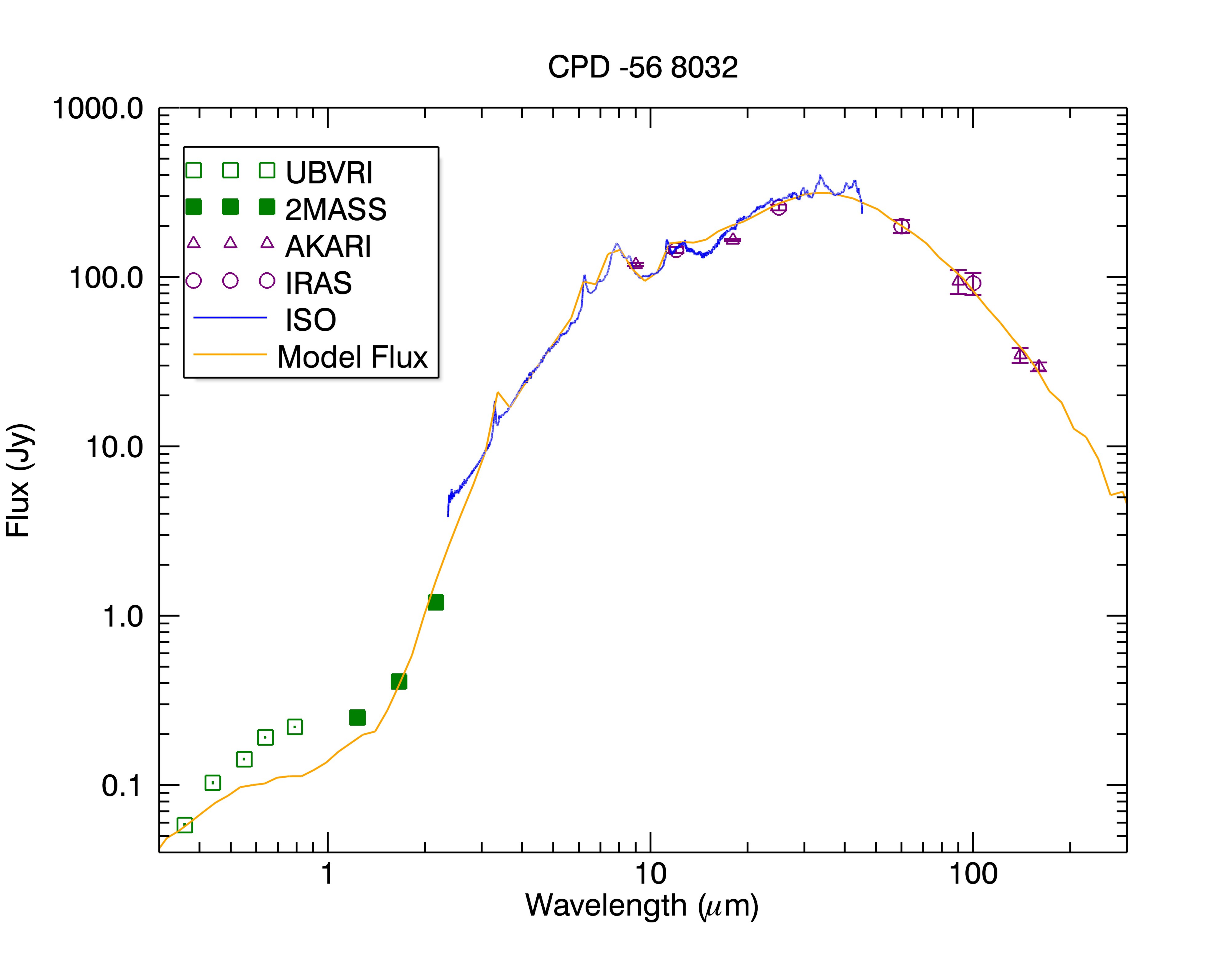}
\end{center}
\caption{SED for CPD --56$^{\rm o}$~8032, and best-fit Monte Carlo RT model. This model consists of amorphous carbon dust in a disk plus a large diffuse envelope. The best fit parameters are shown in Table 4.}
\end{figure}

\begin{figure}
\figurenum{5} 
\begin{center}
\includegraphics[width=5in,angle=0]{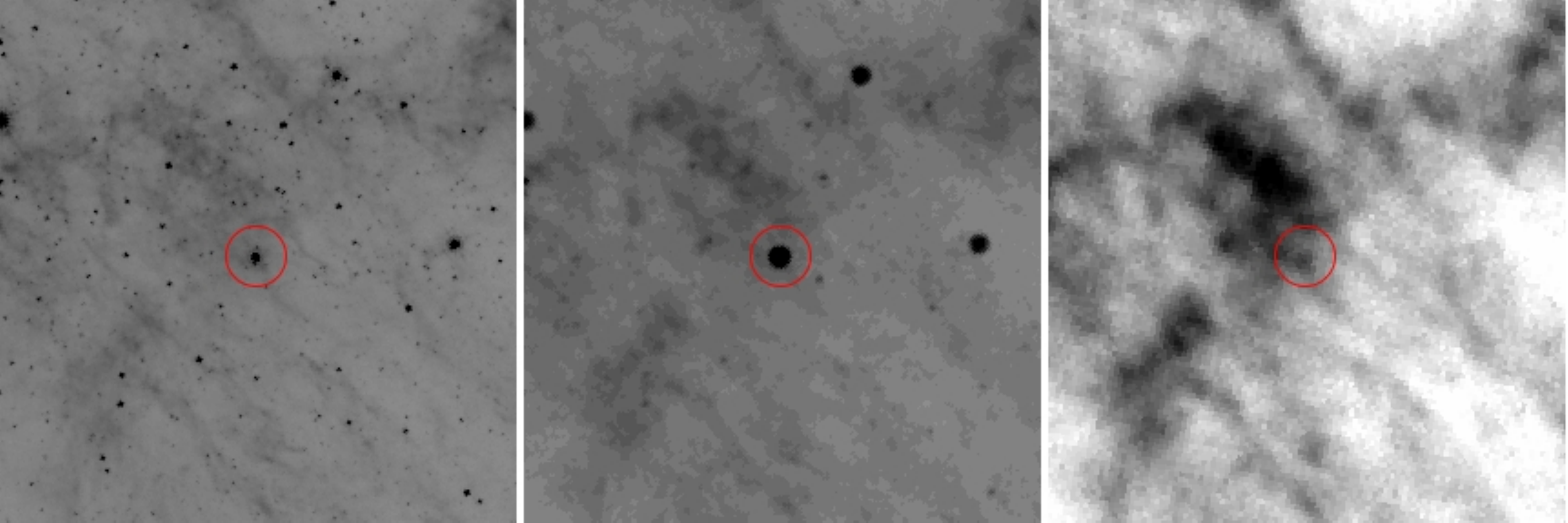}
\end{center}
\caption{Images of HV 2671 in the Spitzer/IRAC 8.0 \micron~(left), Spitzer/MIPS 24 \micron~(center), and SPIRE 250 \micron~(right) bands. The field is $\sim$12\arcmin~x 12\arcmin.}
\end{figure}

\clearpage

\begin{deluxetable}{ll}
\tabletypesize{\scriptsize}
\tablewidth{0pt} 
\tablecaption{HV 2671 Photometry}
\tablenum{1}
\tablehead{\colhead{Band}&
           \colhead{Flux (Jy)$^a$}}
\startdata
B&1.24e-03 $\pm$ 1.10e-05\\
V&1.49e-03 $\pm$ 1.40e-05\\
I$_c$&1.39e-03 $\pm$ 1.30e-05\\
2MASS/J  & 1.36e-03   $\pm$   4.50e-05\\
2MASS/H &  1.58e-03    $\pm$  6.20e-05\\
2MASS/K  & 3.41e-03   $\pm$   9.20e-05\\
IRAC/3.6 &  2.78e-02   $\pm$   5.28e-04\\
IRAC/4.5 &  4.97e-02   $\pm$   1.18e-03\\
IRAC/5.8  & 7.42e-02   $\pm$   1.39e-03\\
IRAC/8.0 &  9.91e-02   $\pm$   1.30e-03\\
MSX/A(8.28)&1.05e-01$\pm$ 6.70e-03\\
AKARI/9&1.06e-01 $\pm$ 7.23e-03 	\\
AKARI/18&1.61e-01 $\pm$ 2.95e-02\\
MIPS/24 &  1.36e-01   $\pm$   6.77e-04\\
IRAS/25&2.09e-01 $\pm$ 1.88e-02\\
PACS/100&4.86e-02 $\pm$5.71e-03\\
PACS/160 & 3.23e-02   $\pm$  4.12e-03\\
SPIRE/250  & 7.92e-02   $\pm$ 6.21e-03\\
\enddata
\tablenotetext{a}{BVI photometry from \citet{2005MNRAS.361.1055S} }
\end{deluxetable}

\begin{deluxetable}{ll}
\tabletypesize{\scriptsize}
\tablewidth{0pt} 
\tablecaption{V348 Sgr Photometry}
\tablenum{2}
\tablehead{\colhead{Band}&
           \colhead{Flux (Jy)$^a$}}
\startdata
U& 2.34e-02  $\pm$ 1.0e-03\\
B&3.39e-02  $\pm$ 1.0e-03\\
V&4.78e-02 $\pm$  1.0e-03\\
R$_c$&5.11e-02  $\pm$ 1.0e-03\\
I$_c$&5.63e-02 $\pm$  1.0e-03\\
2MASS/J  & 1.09e-01 $\pm$  2.0e-03\\
2MASS/H &  2.71e-01 $\pm$  6.0e-03\\
2MASS/K  & 7.73e-01  $\pm$ 1.5e-02\\
AKARI/18&2.81e+00  $\pm$     2.8e-02\\
IRAS/12&5.53e+00    $\pm$   2.2e-01\\
IRAS/25&3.00e+00   $\pm$    2.4e-01\\
IRAS/60&2.88e+00	$\pm$   2.9e-01\\
AKARI/90&1.64e+00 $\pm$1.8e-02\\
\enddata
\tablenotetext{a}{UBVRI photometry from  \citet{1985AAS...61..375H}}
\end{deluxetable}

\begin{deluxetable}{ll}
\tabletypesize{\scriptsize}
\tablewidth{0pt} 
\tablecaption{CPD --56$^{\rm o}$~8032  Photometry}
\tablenum{3}
\tablehead{\colhead{Band}&
           \colhead{Flux (Jy)$^a$}}
\startdata
U& 5.80e-02 $\pm$ 1.0e-03\\
B&1.03e-01 $\pm$ 1.0e-03\\
V&1.42e-01 $\pm$ 1.0e-03\\
R$_c$&1.91e-01 $\pm$ 1.0e-03\\
I$_c$&2.20e-01 $\pm$ 1.0e-03\\
2MASS/J  & 2.50e-01$\pm$  6.0e-02\\
2MASS/H & 4.08e-01 $\pm$ 1.0e-02\\
2MASS/K  & 1.20e+00$\pm$  3.0e-02\\
AKARI/9&1.183e+02 $\pm$ 2.8e+00\\
IRAS/12&1.44e+02 $\pm$  5.8e+00\\
AKARI/18&1.650e+02 $\pm$ 2.3e+00\\
IRAS/25&2.57e+02  $\pm$ 1.0e+01\\
IRAS/60&1.99e+02  $\pm$ 1.8e+01\\
IRAS/100&9.17e+01 $\pm$  1.4e+01\\
AKARI/90&9.441e+01  $\pm$ 1.51e+01\\
AKARI/140&3.459e+01 $\pm$  3.46e+00\\
AKARI/160&2.943e+01 $\pm$  1.77e+00\\
\enddata
\tablenotetext{a}{UBVRI photometry from  \citet{1999Obs...119...76J}}
\end{deluxetable}

\begin{deluxetable}{lllcccccc}
\tabletypesize{\scriptsize}
\tablewidth{0pt} 
\tablecaption{Stellar and Dust Parameters}
\tablenum{4}
\tablehead{\colhead{Name}&
           \colhead{T$_{eff}^a$}&
           \colhead{L$_{Bol}^b$}&
           \colhead{d$^a$}&
           \colhead{Disk Mass}&
          \colhead{Disk Radius$^c$} &
          \colhead{Envelope Mass}&
          \colhead{Envelope Radius$^c$}&
          \colhead{$i$}
           \\
           \colhead{}&
           \colhead{(K)}&
           \colhead{(L$_{\sun}$)}&
           \colhead{(kpc)}&
           \colhead{(M$_{\sun}$)}&        
           \colhead{(AU)}&
           \colhead{(M$_{\sun}$)}&
             \colhead{(AU)}&
             \colhead{(deg)}
           }
\startdata
HV 2671&20k&10\,900&50&0.07&48/400&5&48/50\,000&65\\
V348 Sgr&20k&6000&5.4&0.05&11/70&1.5$\times$10$^{-5}$&22/400&$<$70\\
CPD--56$^{\rm o}$8032&34.5k&13\,700&1.35&0.08&83/400&1.5&168/23\,000&45\\
\enddata
\tablenotetext{a}{\citep{1997MNRAS.292...86D,2002AJ....123.3387D}}
\tablenotetext{b}{Bolometric luminosity for the best fit RT models plotted in Figures 2-4.}
\tablenotetext{c}{The two numbers represent the inner and outer radii.}
\end{deluxetable}


\end{document}